\documentclass[aps,prb,nofootinbib,twocolumn,amsmath,amssymb,superscriptaddress,notitlepage]{revtex4-2}
\usepackage{latexsym}
\usepackage{graphicx}
\usepackage{times,psfrag,subfigure}
\usepackage{enumerate}
\usepackage{amsmath}
\usepackage{amsthm}
\usepackage{dsfont}
\usepackage{dcolumn}
\usepackage{bm,bbm}
\usepackage{wasysym}
\usepackage[normalem]{ulem} 
\usepackage{color}
\usepackage[dvipsnames]{xcolor}
\usepackage[breaklinks,colorlinks,allcolors=blue]{hyperref}
\usepackage{latexsym,amsmath,amssymb,bm,euscript}
\usepackage{textcomp}
\usepackage{tabularx}
\usepackage{multirow}
\usepackage{setspace}
\usepackage{ctable}
\usepackage{sidecap}
\usepackage{placeins}
\usepackage{threeparttable}
\usepackage{braket}
\usepackage{multirow}
\usepackage{comment}
\usepackage{float}
\usepackage{framed}
\usepackage{scalerel,stackengine}
\stackMath

\DeclareMathOperator{\tr}{Tr}
\usepackage{ulem}

\let \oldbm \bm
\renewcommand{\vec}[1]{\oldbm{#1}}

\def\H{\mathcal{H}}
\def\T{\mathcal{T}}
\def\K{\mathcal{K}}
\def\L{\mathcal{L}}

\def\P{\mathcal{P}}

\newcommand{\sket}[1]{|#1\rangle\!\rangle}
\newcommand{\sbra}[1]{\langle\!\langle #1 |}
\newcommand{\sbraket}[1]{\langle\!\langle #1 | #1 \rangle\!\rangle }

\newcommand\reallywidehat[1]{%
\savestack{\tmpbox}{\stretchto{%
  \scaleto{%
    \scalerel*[\widthof{\ensuremath{#1}}]{\kern-.6pt\bigwedge\kern-.6pt}%
    {\rule[-\textheight/2]{1ex}{\textheight}}
  }{\textheight}%
}{0.5ex}}%
\stackon[1pt]{#1}{\tmpbox}%
}

\makeatletter
\renewcommand*\env@matrix[1][*\c@MaxMatrixCols c]{%
  \hskip -\arraycolsep
  \let\@ifnextchar\new@ifnextchar
  \array{#1}}
  
\makeatother

\allowdisplaybreaks

\graphicspath{{figures/}}
\begin{document}

\title{Simulation of bilayer Hamiltonians based on monitored quantum trajectories
}

\author{Yuan Xue}
\affiliation{Department of Physics, The University of Texas at Austin, Austin, Texas 78712, USA}

\author{Zihan Cheng}
\affiliation{Department of Physics, The University of Texas at Austin, Austin, Texas 78712, USA}
\affiliation{Department of Physics, National University of Singapore, Singapore 117551, Singapore}

\author{Matteo Ippoliti}
\affiliation{Department of Physics, The University of Texas at Austin, Austin, Texas 78712, USA}

\date{\today}

\begin{abstract}
    In the study of open quantum systems it is often useful to treat mixed states as pure states of a fictitious doubled system. In this work we explore the opposite approach: mapping isolated bilayer systems to open monolayer systems. Specifically, we show that arbitrary bilayer Hamiltonians possessing an antiunitary layer exchange symmetry, and subject to a constraint on the sign of interlayer couplings, can be mapped to Lindbladians on a monolayer system with some of the jump operators postselected on a fixed outcome (``monitored''). Low-energy states of the bilayer Hamiltonian then correspond to late-time states of the monolayer dynamics. Simulating the latter by quantum trajectory methods has the potential of substantially reducing the computational cost of estimating low-energy observables in the bilayer Hamiltonian by effectively halving the system size. The overhead due to sampling quantum trajectories can be controlled by a suitable importance sampling scheme. We show that, when the quantum trajectories exhibit free fermion dynamics, our approach reduces to the auxiliary field quantum Monte Carlo (AFQMC) method. This provides a physically transparent interpretation of the AFQMC sign-free criteria in terms of properties of quantum dynamics. 
    Finally, we benchmark our approach on the 1D quantum Ashkin–Teller model.
\end{abstract}

\maketitle

\tableofcontents

\section{Introduction}
Recent progress in the study of mixed state phases of matter~\cite{PRXQuantum.4.030317,Lee2025symmetryprotected,bao2023mixedstatetopologicalordererrorfield,PRXQuantum.5.020343,PhysRevB.110.155150,PRXQuantum.6.010344} has opened new avenues for understanding many-body physics in open quantum systems. A central tool in this development is the Choi–Jamiołkowski isomorphism \cite{CHOI1975285,JAMIOLKOWSKI1972275}, which maps a density matrix $\rho$ to a pure state 
$\sket{\rho}$ in a doubled Hilbert space, and thus allows the extension of pure-state concepts and techniques to the mixed-state case. This framework has proven fruitful in characterizing various quantum phases of matter, including mixed symmetry-protected topological (SPT) phases \cite{PhysRevX.13.031016,PhysRevX.15.021062,PRXQuantum.6.010348,Lee2025symmetryprotected} and mixed topological order (TO) \cite{PRXQuantum.6.010313,PRXQuantum.6.010314,PRXQuantum.6.010315,PhysRevB.111.115137}, providing a unified perspective on mixed-state quantum phases.

In this work we pursue a complementary perspective, by understanding certain pure-state ``bilayer'' systems within a mixed-state framework. 
We focus on bilayer quantum systems whose Hamiltonian $\mathcal{H}$ meets two criteria:
(i) it is invariant under the composition of layer exchange and an antiunitary transformation (such as time reversal or particle-hole), which we call ``antiunitary layer exchange'' in the following, 
and (ii) the interlayer couplings obey a sign constraint to be specified below.
Subject to these constraints, we show that the thermodynamics of the bilayer Hamiltonian (controlled by the imaginary time evolution $e^{-\beta \mathcal{H}}$) is related to the {\it dynamics} of an open and monitored monolayer system (controlled by real-time evolution $e^{t\L}$ under some generator $\L$). 
To make this equivalence precise, we associate to each $\mathcal{H}$ obeying the criteria above a Lindbladian $\mathcal{L}$ whose {jump operators} are related to the Hamiltonian interaction terms; furthermore, some of the jumps may be \textit{postselected} onto the ``no-click'' condition to faithfully emulate the target bilayer Hamiltonian.

This mapping has several implications. First, in terms of physics, it relates phase transitions in bilayer systems to dynamical transitions in open monolayers. (See also similar observations in prior works \cite{lu2024bilayerconstructionmixedstate, PhysRevB.110.L241105, PhysRevB.111.054106, PRXQuantum.5.020332}.)
Additionally, it has implications for the complexity of numerically simulating low-energy properties of bilayer quantum Hamiltonians, which are crucial in many areas of condensed matter physics \cite{PhysRevB.49.1397,PhysRevLett.85.1524,PhysRevB.79.220504,PhysRevLett.110.216405,PhysRevX.5.041041,doi:10.1126/science.aal5304,PhysRevLett.69.2863,RevModPhys.77.259,annurev:/content/journals/10.1146/annurev-conmatphys-020911-125018}.
A bilayer system that comprises a total of $2N$ sites ($N$ per layer) has a Hilbert space of dimension $2^{2N}$ (we assume qubit or fermion degrees of freedom throughout the paper). 
A mixed state of an $N$-site monolayer is represented by a density matrix of size $2^N \times 2^N$---the same amount of information. However, the monolayer dynamics may be decomposed into \emph{quantum trajectories} \cite{PhysRevLett.68.580,PhysRevA.45.4879,Molmer:93,RevModPhys.70.101,Daley04032014, PhysRevA.46.4363}, an ensemble of stochastically-sampled pure-state wavefunctions, thus reducing the problem size to $2^N$. This is achieved at the expense of collecting many statistical samples of quantum trajectories to be averaged over. 
Provided the sampling overhead is controlled, this represents a quadratic improvement in complexity that may increase the finite-size limit in numerical simulations by up to a factor of 2.

An important subtlety with this method is the presence of postselected quantum jump operators in the monolayer's dynamics, which makes the sampling of quantum trajectories (for the non-postselected jumps) nontrivial; a na\"ive approach incurs an exponential cost. To avoid this issue, we implement an \textit{importance sampling} process over quantum trajectories tailored to two different types of operators of interest in the bilayer system: intralayer operators (supported in only one of the layers), and symmetric interlayer operators (supported in both layers and obeying the antiunitary layer exchange symmetry). 
We show that, with our importance sampling approach, the variance of Monte Carlo estimates for these types of operators is bounded by constants (in system size). 

The quantum trajectory approach is most powerful when the trajectories themselves admit efficient classical simulation, as is the case when they describe non-interacting fermions. Then the complexity drops from exponential to polynomial in system size $N$. 
We show that in this case, our approach reduces to the
auxiliary-field quantum Monte Carlo (AFQMC) algorithm \cite{PhysRevD.24.2278,PhysRevB.55.7464,PhysRevLett.90.136401,lee2022yearsauxiliaryfieldquantummonte}.
As suggested by the name, AFQMC is based on using an auxiliary field to decouple the density-density interaction into quadratic fermionic terms, via a Hubbard-Stratonovich transformation, with the desired results obtained as averages over multiple realizations of the (randomly sampled) auxiliary field. Within our framework, this auxiliary field can be interpreted as describing different realizations of the dissipation in the open system. 
Notably, the well-known criteria that guarantee absence of the ``sign problem'' in AFQMC acquire a transparent physical meaning in our mapping: they reduce to the requirements of a physical density matrix evolution for the monolayer system.  

The rest of the paper is laid out as follows. In Sec. \ref{sec: Setup}, we introduce the mapping between bilayer Hamiltonians and monolayer dynamics. 
In Sec.~\ref{sec: quantum trajectory} we discuss the quantum trajectory simulation of partially postselected Lindbladians and the importance sampling method.
In Sec. \ref{sec: afqmc}, we clarify the connections between AFQMC and our open system approach, providing physical interpretations of the numerical technique. 
In Sec. \ref{sec: examples}, we benchmark our method against exact numerical results using the one-dimensional quantum Ashkin-Teller model as a test case. 
Finally we discuss our results and open questions in Sec.~\ref{sec:discussion}. 

\section{Mapping bilayer Hamiltonians to open monolayer dynamics \label{sec: Setup}}

\begin{figure}
    \centering
    \includegraphics[width=\columnwidth]{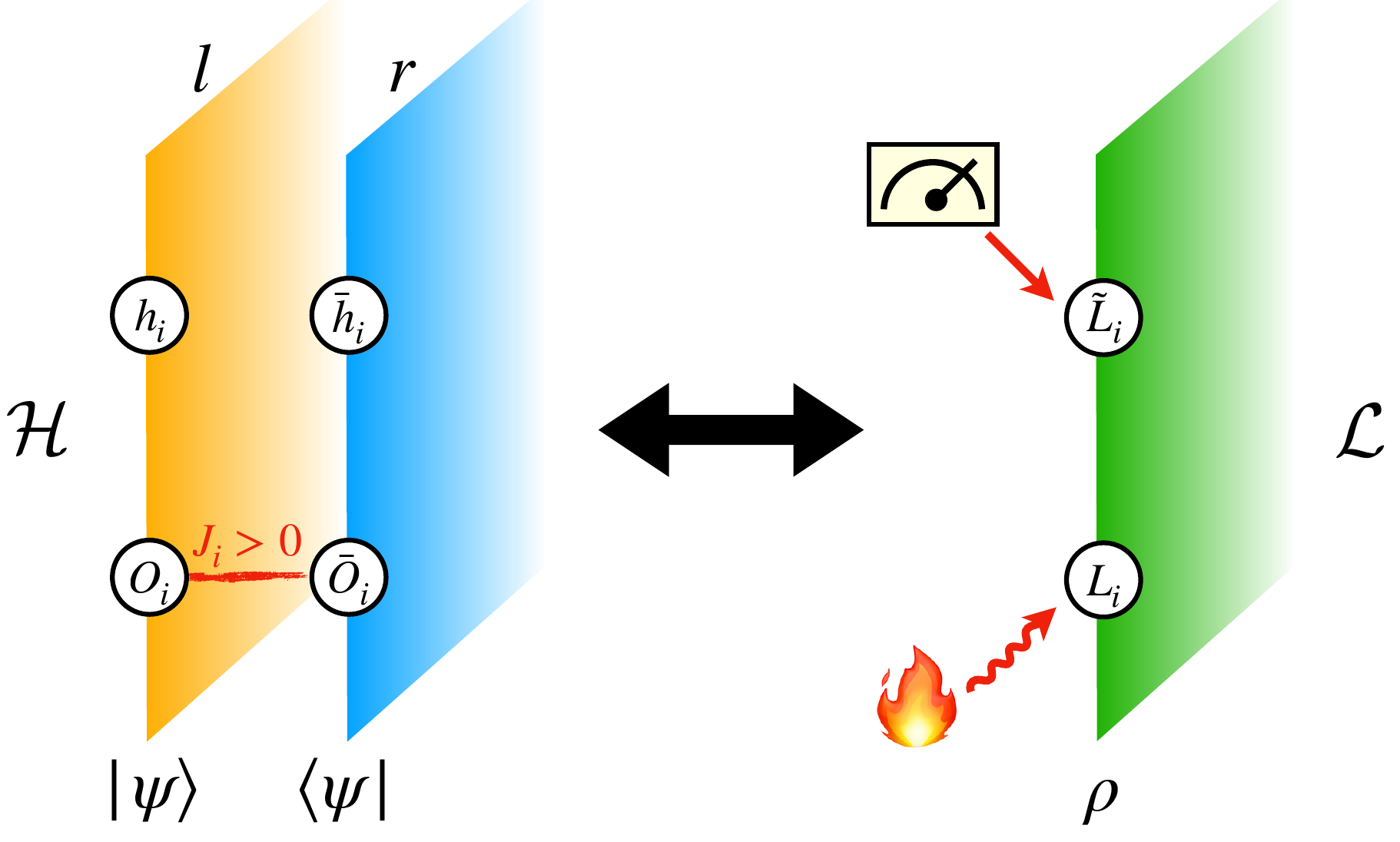}
    \caption{Schematic of the mapping between bilayer Hamiltonians (left) and monolayer dynamics (right). The two layers $l$, $r$ are viewed as the ``ket'' and ``bra'' sides of a density matrix $\rho$. The bilayer Hamiltonian $\H$ comprises intralayer terms $h_{i,l}$, $\bar{h}_{i,r}$ and interlayer couplings $J_i O_{i,l} \bar{O}_{i,r}$, with $J_i > 0$ and the bar denoting an antiunitary transformation. The monolayer dynamics $\L$ comprises dissipation (jump operators $L_i$) and monitoring (jump operators $\Tilde{L}_i$), see Eq.~(\ref{eq:Li_def},\ref{eq:tildeLi_def}). }
    \label{fig:idea}
\end{figure}

In this Section we present the theoretical framework for our mapping between bilayer Hamiltonians and open/monitored monolayer dynamics, schematically illustrated in Fig.~\ref{fig:idea}. 
Our mapping is based on the standard identification between the space of linear operators on a vector space $V$, ${\rm End}(V)$, and the tensor product $V\otimes V^\ast$ ($V^\ast$ is the dual vector space). In quantum physics language, this corresponds to the mapping $\ket{v} \bra{w} \leftrightarrow \ket{v} \otimes \ket{w}$ between operators and states of a doubled Hilbert space. 
Our mapping can be stated as 
\begin{equation}
    e^{-\beta \H} \propto e^{t\L}, 
    \label{eq: expH_expL}
\end{equation}
where $\H$ is a Hamiltonian on a doubled Hilbert space, representing a bilayer quantum system, and $\L$ is a generator of dynamics for a single Hilbert space. Thus thermal states of the bilayer system at inverse temperature $\beta$ are related to dynamics of the monolayer system for time $t = \beta$. Below we present the assumptions on $\H$ and the details of the mapping. 

We focus on bilayer Hamiltonians invariant under the composition of layer exchange and an antiunitary symmetry (``antiunitary layer exchange'' for short)
\begin{equation}
    \H = H_l + \bar{H}_r - \sum_i J_i O_{l,i}\bar{O}_{r,i}, \qquad J_i \geq 0.
    \label{eq: setup H}
 \end{equation}
Here $l$ (``left'') and $r$ (``right'') label the two layers, $H_l=H\otimes I$ and $\bar{H}_r = I\otimes \bar{H}$ are intralayer Hamiltonians, and the remaining terms describe interlayer couplings, with the $O_i$ being local operators. 
The overline represents an arbitrary antiunitary transformation; different choices are related by a unitary on layer $r$. Conventional choices for the antiunitary transformation are:
\begin{itemize}
    \item For qubit systems, the time reversal transformation $\T = \K \bigotimes_j (i Y_j)$, with $\K$ the complex conjugation, which acts as $\T \boldsymbol{\sigma}_j \T^{-1} = - \boldsymbol{\sigma}_j$ on spin operators;
    \item For fermion systems, the particle-hole transformation $\P$ defined by $\P c_j \P^{-1} = c_j^\dagger$, $\P c_j^\dagger \P^{-1} = c_j$, or the time reversal transformation $\T$ defined by $\T c_j \T^{-1} = c_j$, $\T c_j^\dagger \T^{-1} = c_j^\dagger$ (the two are related by the unitary $\P \T$ transformation). 
\end{itemize}
One can see that Eq.~\eqref{eq: setup H} is invariant under the combination of layer exchange $l\leftrightarrow r$ and the antiunitary transformation $\mathcal{H} \leftrightarrow \bar{\mathcal H}$. 
Bilayer Hamiltonians with such a layer exchange symmetry are physically relevant in a variety of condensed matter settings, from spin ladders to Hubbard models to moir\'e materials (some of these examples are discussed in Sec.~\ref{sec: examples}).

To realize the mappping in Eq.~\eqref{eq: expH_expL}, we aim to identify the bilayer Hamiltonian $\H$ [Eq.~\eqref{eq: setup H}] with a suitable dynamical generator $\L$ of the form
\begin{equation}
    \L[\rho] = \sum_i L_i\rho L^\dagger_i  
    - \frac{1}{2} \{ L_i^\dagger L_i, \rho \} - \frac{1}{2} \{\Tilde{L}_i^\dagger \Tilde{L}_i,\rho\} ,
    \label{eq: lindbladian}
\end{equation}
where $\{ L_i\} $ and $\{ \Tilde{L}_i\}$ are two families\footnote{Note the number of operators $\{L_i\}$ and $\{\Tilde{L}_i\}$ could be different; if so we pad one of the lists with zeros.} of ``jump operators''.
As defined, $\L$ would be a Lindbladian~\cite{10.1063/1.522979,Lindblad1976,PhysRevA.89.022118} except for the absence of $\Tilde{L}_i \rho \Tilde{L}_i^\dagger$ terms, representing {\it postselection} of the $\tilde{L}_i$ jumps on the ``no-click'' condition. 
This makes the dynamical map $e^{t\L}$ completely positive but generally not trace-preserving. (The loss of trace is related to the decaying probability of successful postselection.) 
As a consequence, the dynamics generated by $\L$ is both {\it open} (due to the $L_i$ jumps which induce decoherence) and {\it monitored} (due to the postselected $\Tilde{L}_i$ jumps which effectively implement continuous measurement). 
Note also that the standard Lindblad master equation admits a coherent contribution $-i[H,\rho]$ that we take to be zero. Such a contribution would make $\L$ non-Hermitian and break the mapping to a standard (Hermitian) Hamiltonian $\H$. 

To identify $\H$ [Eq.~\eqref{eq: setup H}] and $\L$ [Eq.~\eqref{eq: lindbladian}] we use the the Choi–Jamiołkowski isomorphism \cite{CHOI1975285,JAMIOLKOWSKI1972275} in the form $A\rho B \mapsto (A\otimes \bar{B})\sket{\rho}$ [notice the antiunitary transformation of $B$, necessary to consistently map dual vectors (``bras'') to vectors (``kets'')].
The desired mapping is obtained by setting
\begin{align}
    L_i & = \sqrt{J_i} O_i, \label{eq:Li_def} \\ 
    \Tilde{L}_i & = \sqrt{\kappa_i + 2h_i - L_i^\dagger L_i} \label{eq:tildeLi_def}
\end{align}
with $h_i$ a local interaction term in the intralayer Hamiltonian $H$, such that $\sum_i h_i = H$, and $\kappa_i$ a constant large enough to make $\kappa_i + 2h_i - L_i^\dagger L_i \geq 0$, so that the operator square root is well-defined. 
Notice that Eq.~\eqref{eq:Li_def} necessitates $J_i > 0$. 
One can verify that 
\begin{align} 
\L[\rho] & = \sum_i J_i O_i \rho O_i - \frac{1}{2} \{ \kappa_i + 2h_i, \rho\} \nonumber \\
& = -H\rho - \rho H - \kappa_{\rm tot} \rho + \sum_i J_i O_i \rho O_i  \label{eq:linbladian}
\end{align}
with $\kappa_{\rm tot} = \sum_i \kappa_i$; 
then, ``vectorizing'' Eq.~\eqref{eq: lindbladian} yields
\begin{align}
    \mathcal{L}\sket{\rho} = \left( \sum_i J_i O_i \otimes \bar{O}_i - H \otimes I - I\otimes \bar{H} -\kappa_{\rm tot} \right) \sket{\rho},
\end{align}
where the density matrix $\rho$ was mapped to a pure state $\sket{\rho}$ of the doubled Hilbert space. 
Thus, up to labeling the Hilbert space copies as the layers $l,r$ we conclude that $\L = -\kappa_{\rm tot} - \H$. This gives
\begin{equation}
    e^{-\beta \H} = e^{\kappa_{\rm tot} \beta} e^{\beta \L}
\end{equation}
which is Eq.~\eqref{eq: expH_expL} up to identifying $t = \beta$. 

Some remarks are in order. 
First, in Eq. \eqref{eq: expH_expL}, we identify the inverse temperature $\beta=1/T$ of the bilayer system with the time $t$ in the monolayer dynamics. The finite-time state given by $\rho_t\propto e^{t\L}[I]$ corresponds to a state cooled to temperature $T = 1/t$: longer time evolution corresponds to lower temperature, and infinite time (steady state) corresponds to the ground state. So we anticipate quantum phase transitions in $\H$ to map onto steady-state transitions in $\L$, and thermal transitions in $\H$ to map onto dynamical (finite-time) transitions in $\L$. 

Secondly we note that, if $H = \frac{1}{2} \sum_i L_i^\dagger L_i$, then we can choose $2h_i = L_i^\dagger L_i$ and thus set $\kappa_i = \Tilde{L}_i = 0$. In this case $\L$ reduces to a standard Lindbladian without postselection~\cite{pjs3-14cc}.
This condition however drastically restricts the form of allowed intralayer interactions $H$ in the bilayer Hamiltonian $\H$. Specifically, intralayer interactions are ``locked'' to the interlayer interactions via $H = \frac{1}{2} \sum_i J_i O_i^2$. The freedom to independently tune intralayer and interlayer interactions comes at the expense of postselection. The probability of no clicks during a small time interval $dt$ is to leading order $1 - dt \sum_i \tr (\rho \Tilde{L}_i^\dagger \Tilde{L}_i)$, so the decay rate for the probability of no clicks is set by the size of $\Tilde{L}_i$, i.e., the discrepancy between interlayer and intralayer interactions. Note also that the probability of success decays exponentially in system size (due to the sum over jumps $\tilde{L}_i$, assumed to occur everywhere in the system) for a fixed evolution time. Effectively this limits the evolution time and system size achievable in experiments, and thus the temperature and system size that can be simulated in the virtual bilayer system.

Lastly, we remark on how observables on the virtual bilayer system can be accessed from the monolayer density matrix. 
Let us consider the expectation values of tensor-product observables $A_l \bar{B}_r$ in the bilayer systems. Identifying the bilayer pure state $\ket{\Psi}_{lr}$ with a vectorized density matrix $\sket{\rho_t}$,
one obtains a R\'enyi-2 correlator:
\begin{equation}
    {}_{lr}\!\bra{\Psi} A_l \bar{B}_r \ket{\Psi}_{lr}  
    = \frac{ \sbra{\rho_t} A\otimes \bar{B} \sket{\rho_t} }{\sbraket{\rho_t}} 
    = \frac{\tr(\rho_t A \rho_t {B})}{\tr\rho^2_t}.
    \label{eq: expectation_observable}
\end{equation}
Therefore, long-range order in exciton-like correlators (of the form $\langle A_l \bar{A}_r\rangle$) in the bilayer system maps onto long-range order in R\'enyi-2 correlators for the monolayer density matrix. 
This is closely related to the phenomenon of {\it spontaneous strong-to-weak symmetry breaking}\footnote{Note however that R\'enyi-2 correlators are only a qualitative proxy for SWSSB as they are not stable under strongly-symmetric channels~\cite{PRXQuantum.6.010344}.} (SWSSB)~\cite{PRXQuantum.6.010344,feng2025hardness, PhysRevB.110.155150,PhysRevB.111.115137, PhysRevLett.134.150405}. 

To summarize, our mapping shows that open and monitored quantum dynamics, featuring quantum jumps some of which may be postselected on no-clicks, can emulate the physics of a wide class of bilayer quantum Hamiltonians [Eq.~\eqref{eq: setup H}], which encompasses many interesting examples including Hubbard models. 
The two constraints on the bilayer Hamiltonian have a clear physical meaning:
the antiunitary layer exchange is a consequence of the fact that the two layers emerge from the ``ket'' and ``bra'' Hilbert spaces of a density matrix, whose time evolution is not independent but instead related by time-reversal;
the sign constraint on interlayer couplings reflects their origin as noise processes (with a real, positive rate). 
While the mapping of Lindbladians to bilayer Hamiltonians is a standard tool, the inclusion of postselected jumps allows us to achieve a much wider and more interesting class of bilayer Hamiltonians.

\section{Quantum trajectories \label{sec: quantum trajectory}}

Besides the physical insight, an interesting consequence of our method is in numerical simulation. 
Indeed, bilayer systems with $N$ qubits per layer have a full Hilbert space of dimension of $2^{2N}$. In contrast, our approach operates entirely within a monolayer system, on a Hilbert space of dimension $2^N$. Of course, since the monolayer system is open, it is described by a $2^N \times 2^N$-dimensional density matrix, giving again $4^N$ variables; but a substantial reduction in computational complexity is possible by making use of {\it quantum trajectory} methods, where the density matrix is unraveled into an ensemble of stochastically-sampled pure states \cite{PhysRevLett.68.580,PhysRevA.45.4879,Molmer:93,RevModPhys.70.101,Daley04032014, PhysRevA.46.4363}. 
This reduction in Hilbert space dimension comes with a trade-off in the form of averaging over quantum trajectories, with a large number of samples needed to achieve accurate results. However we will show that this sampling overhead can be bounded by a constant in many cases of practical interest, so the reduction in Hilbert space dimension is advantageous (at least asymptotically for large systems).

We begin with a brief review of the quantum trajectory method for open quantum systems in general (Sec.~\ref{subsec: trajectory method}), then discuss the required modifications when the underlying dynamics includes postselection (Sec.~\ref{subsec: postselection}). 
We then introduce, through an instructive minimal example, the application of quantum trajectories to bilayer systems via our mapping (Sec.~\ref{subsec: dimer example}). Finally, we incorporate \emph{importance sampling} in the quantum trajectory method and show that the variance of observable estimators can be bounded (Sec.~\ref{subsec: importance}).

\subsection{Review \label{subsec: trajectory method}}

The ``quantum trajectory'' or ``stochastic wavefunction'' method is widely used to simulate open quantum system dynamics. 
Instead of simulating the evolution of mixed states $\rho(t) = e^{t \L}[\rho(0)]$, one unravels the mixed state into an ensemble of pure quantum trajectories whose statistical average yields the evolution of the
mixed state. 
Consider as an example the following stochastic Schr\"odinger equation for a pure state:
\begin{equation}
    d \ket{\psi} = -H_\text{eff}\ket{\psi} dt + i\sum_i dw_i L_i \ket{\psi},
    \label{eq: unravel pure states}
\end{equation}
where $dw_i$ are random variables with $\overline{dw_i} = 0$ and $\overline{dw_i dw_j^\ast } = dt \delta_{ij}$.
One can verify that under evolution for an infinitesimal time interval $dt$, the state $\ket{\psi}\!\bra{\psi}$ changes to $(\ket{\psi}+d\ket{\psi})(\bra{\psi}+d\bra{\psi})$, which on average gives 
\begin{align}
    \rho \mapsto \rho - dt\{ H_\text{eff},\rho\} + dt \sum_i L_i \rho L^\dagger_i,
    \label{eq: unravel lindbladian}
\end{align}
which is nothing but the Lindblad master equation, if $H_{\rm eff} = \frac{1}{2} \sum_i L_i^\dagger L_i$. (Here, like before, we neglect the coherent part $-i[H,\rho]$ of the Lindbladian equation.)

We may also recover our partially monitored dynamics $\L$, Eq.~\eqref{eq: lindbladian}, by setting $H_{\rm eff} = \frac{1}{2} \sum_i L_i^\dagger L_i + \Tilde{L}_i^\dagger \Tilde{L}_i$; we return to this below, and focus on un-monitored dynamics for now.  
Simulation of an ensemble of pure-state solutions to the stochastic Schr\"odinger equation then gives, on average, the density matrix $\rho$ that solves the master equation, and observable expectation values can be computed as $\mathbb{E}[ \bra{\psi} O \ket{\psi}] = \tr (\rho O)$. 

For the purpose of numerical simulation, it is convenient to integrate the master equation over a small time step $\Delta t$ to get the quantum channel 
\begin{equation}
    e^{\Delta t \L}[\rho] = \sum_{i=0}^{m} K_i \rho K_i^\dagger,
\end{equation}
where $m$ is the number of jump operators $L_i$, and the $K_i$ are Kraus operators
\begin{align}
    K_0 & = I - \frac{\Delta t}{2} \sum_{i=1}^m L_i^\dagger L_i, \label{eq:strong_jump1} \\
    K_i & = \sqrt{\Delta t} L_i \quad (i = 1,\dots m). \label{eq:strong_jump2}
\end{align}
They obey the normalization condition $\sum_{i=0}^m K_i^\dagger K_i = I$.
Then one gets the solution as 
\begin{equation} 
\rho(t) \simeq \sum_{\mathbf s} K_{s_n} \cdots K_{s_1} \rho K_{s_1}^\dagger \cdots K_{s_n}^\dagger,
\end{equation}
where each Kraus operator index $s_1,\dots s_n$ plays the role of $i$ above and is summed from $0$ to $m$, the time $t$ is discretized into $n$ intervals of length $\Delta t$, and the approximation is up to errors $O(\Delta t)$. 
If the initial state $\rho(0)$ is pure (it can be unraveled into pure states if necessary), then one can simulate the dynamics of pure states 
\begin{equation}
    \ket{\psi_{\mathbf s}(t)} = \frac{K_{s_n} \cdots K_{s_1} \ket{\psi(0)} }{\sqrt{p(\mathbf s)}},
\end{equation}
where the string $\mathbf s$ labels quantum trajectories, and $p(\mathbf s) = \bra{\psi(0)} K_{s_n}^\dagger \cdots K_{s_1}^\dagger K_{s_1} \cdots K_{s_n} \ket{\psi(0)}$ is a probability distribution over trajectories (normalization of the probability to 1 follows from the Kraus normalization condition). 

To compute an expectation value $\tr(\rho O)$, one needs to evaluate the trajectory average
\begin{equation}
  \tr(\rho O) = \sum_{\mathbf s} p(\mathbf s) \bra{\psi_{\mathbf s}(t)} O \ket{\psi_{\mathbf s}(t)}. 
\end{equation}
The sum contains exponentially many terms, $\sim (m+1)^n$, so an exact calculation is hard, but it can be efficiently approximated by sampling $\mathbf s$ according to the distribution $p(\mathbf s)$: 
\begin{equation}
  \tr(\rho O) = \mathop{\mathbb{E}}_{\mathbf s\sim p} [ \bra{\psi_{\mathbf s}(t)} O \ket{\psi_{\mathbf s}(t)} ] . 
\end{equation}
Trajectories can be sampled according to the desired distribution by noting the Markov property:
\begin{align}
    p(s_{n+1}|\mathbf s_{1:n}) 
    & = \frac{\bra{\psi(0)} K_{s_{1}}^\dagger \cdots K_{s_{n+1}}^\dagger K_{s_{n+1}} \cdots  K_{s_1} \ket{\psi(0)} }{\bra{\psi(0)} K_{s_1}^\dagger \cdots K_{s_n}^\dagger K_{s_n} \cdots  K_{s_1}\ket{\psi(0)} } \nonumber \\
    & = \bra{\psi_{\mathbf s_{1:n}} (t)} K_{s_{n+1}}^\dagger K_{s_{n+1}} \ket{\psi_{\mathbf s_{1:n}} (t)}. \label{eq:markov_property}
\end{align}
At each time $t$ during the evolution, one can sample the next jump $s \in \{0,\dots m\}$ based on the probability distribution $s \sim \langle K_s^\dagger K_s\rangle_t$ at that time. 

To summarize, the standard prescription for quantum trajectory simulation is:
\begin{enumerate}
    \item Start from a state $\ket{\psi_t}$. Compute the distribution over jumps
    \begin{equation}
        p(s) = \bra{\psi_t} K_s^\dagger K_s \ket{\psi_t}.
    \end{equation}
    \item Sample the next jump $s\sim p(s)$, update the state and normalize:
    \begin{equation}
        \ket{\psi_{t+ \Delta t}} = \frac{ K_s \ket{\psi_t} }{\sqrt{p(s)}}
        \label{eq: weak jump}
    \end{equation}
    \item Repeat steps 1 and 2 until the target evolution time is reached, then compute desired observables.
    \item Repeat steps 1, 2, 3 with an independent random realization of the trajectory $\mathbf s$ each time, average the results. 
\end{enumerate}

The choice of Kraus operators is not unique. Eq.~\eqref{eq:strong_jump1}, \eqref{eq:strong_jump2} represent ``strong jumps'', where the state evolves gently most of the time (via $K_0$, the ``no click'' evolution) and occasionally undergoes a strong discontinuous update (via $K_i$, $i = 1,\dots m$). 
Another choice which may lead to lower variation across trajectories is to consider {\it weak jumps}, where all the Kraus operators are close to the identity. 
A possible choice is\footnote{The prefactors $\pm \sqrt{dt}$ in the exponent play an analogous role to the Gaussian variables $dw$ in Eq.~\eqref{eq: unravel pure states}, having zero mean and variance $dt$.}
\begin{equation}
    K_{i,\pm} = \frac{1}{\sqrt{2m}} e^{\pm i \sqrt{\Delta t} L_i  + \Delta t (L_i^2 - L_i^\dagger L_i)/2}.
\end{equation}
In the case of Hermitian jumps, $L_i = L_i^\dagger$, this simplifies to $K_{i,\pm} \propto e^{\pm i \sqrt{\Delta t} L_i}$. These are proportional to unitary operators, and represent dephasing under a weak, temporally-random Hamiltonian evolution.

\subsection{Quantum trajectories with postselection} \label{subsec: postselection}

So far we have reviewed the formalism of quantum trajectories for the standard case of Lindbladian dynamics, without measurements. However, our mapping presented in Sec.~\ref{sec: Setup} generally features monitoring in the form of jumps that are postselected on the ``no-click'' condition. 
This means that, out of the Kraus operators in Eq.~\eqref{eq:strong_jump2}, only some outcomes are allowed, while others are forbidden. 
It is helpful to reintroduce the notation of Sec.~\ref{sec: Setup}: we use $L_i$ for allowed jumps and $\Tilde{L}_i$ for forbidden (monitored) jumps, and thus rewrite 
\begin{equation}
    K_0 = I - \frac{\Delta t}{2} \left( \sum_i L_i^\dagger L_i + \Tilde{L}_i^\dagger \Tilde{L}_i\right)
\end{equation}
and $K_i = \sqrt{\Delta t} L_i$ ($i = 1,\dots m$). Additional Kraus operators $\Tilde{K}_i = \sqrt{\Delta t} \Tilde{L}_i$ do not occur due to postselection. 
This gives 
\begin{equation}
    \sum_{i=0}^m K_i \rho K_i^\dagger = \rho + \Delta t \L[\rho] +O(\Delta t^2)
\end{equation}
with $\L$ as in Eq.~\eqref{eq: lindbladian}. 

This monitoring and postselection of jumps has important consequences on the sampling of trajectories. 
First, removing some Kraus operators breaks the normalization condition:
\begin{align}
    \sum_{i=0}^m K_i^\dagger K_i 
    & = I - \Delta t \sum_i \Tilde{L}_i^\dagger \Tilde{L}_i + O(\Delta t^2) \nonumber \\
    & = I - \sum_i \Tilde{K}_i^\dagger \Tilde{K}_i \leq I.
\end{align}
This causes a decrease in the trace of $\rho$ over time. Physically this represents the decay in the probability of obtaining a trajectory where the $\Tilde{L}_i$ jumps do not occur. 
As a consequence, to compute observable expectation values, one must explicitly normalize $\rho$ by its trace:
\begin{align}
    \frac{\tr(\rho O)}{\tr(\rho)} 
    & = \frac{\sum_{\mathbf s \in \mathcal{A}}  p(\mathbf s) \bra{\psi_{\mathbf s}} O \ket{\psi_{\mathbf s}} }{\sum_{\mathbf s \in \mathcal{A}} p(\mathbf s) } \nonumber \\
    & = \sum_{\mathbf s \in \mathcal{A}}  p(\mathbf s | \mathcal{A}) \bra{\psi_{\mathbf s}} O \ket{\psi_{\mathbf s}} \label{eq:conditional_probability}
\end{align}
where $\mathcal{A}$ represents the set of allowed trajectories, those in which the monitored jumps do not occur. We also introduced $p(\mathbf{s} | \mathcal{A}) = p(\mathbf s) / \sum_{\mathbf s'\in \mathcal{A}} p(\mathbf s')$, the {probability} of trajectory $\mathbf s$ conditional on the monitored jumps not happening. 

Instead of working with an incomplete set of Kraus operators, it will be more convenient to view the dynamics as a combination of noise, with a complete set of Kraus operators [for example those in Eq.~(\ref{eq:strong_jump1},~\ref{eq:strong_jump2})], and imaginary time evolution under an ``effective Hamiltonian''
\begin{equation}
    H_{\rm eff} = \frac{1}{2} \sum_i \Tilde{L}_i^\dagger \Tilde{L}_i. 
\end{equation}
Trajectories then are given by 
\begin{equation}
    \ket{\psi_{\mathbf s}} \propto e^{-\Delta t H_{\rm eff}} K_{s_n}  \cdots e^{-\Delta t H_{\rm eff}}  K_{s_1} \ket{\psi(0)},
    \label{eq:trajectories_def}
\end{equation}
with $\mathbf{s}$ ranging only over the set $\mathcal{A}$ of allowed trajectories.

Crucially, the conditional probability $p(\mathbf s|\mathcal{A})$ does not obey the Markov property, Eq.~\eqref{eq:markov_property}. The standard procedure of sampling the next jump at each step in the time evolution therefore does not reproduce the correct trajectory distribution. It is necessary to sample trajectories according to their global history. We clarify this aspect with a simple example next. 

\subsection{Minimal example: Ising dimer} \label{subsec: dimer example}

Consider the minimal example of a ``bilayer'' Hamiltonian: a dimer, with two qubits labeled $l$ and $r$ respectively. We take the Hamiltonian 
\begin{equation}
    \H = J X_l X_r + h Z_l - h Z_r,
\end{equation}
which is of the form in Eq.~\eqref{eq: setup H} with $O_i = X$, $J>0$, and $H = hZ$. 
The antiunitary transformation is time reversal, so $\bar{Z} = -Z$. 
We take the initial state $\ket{0}_l \ket{0}_r$ and cool it via imaginary time evolution, $e^{-\beta \H} \ket{0}_l \ket{0}_r$. We aim to evaluate the expectation value $\langle Z_l\rangle$ on this state of the dimer. 

Under our mapping, the cooled state of the dimer is equivalent to a time-evolved state of a single qubit: $e^{\beta \L}[\ket{0}\!\bra{0}]$, with $\L$ given by
\begin{equation}
    \L[\rho] = J(X\rho X - \rho) - \{ hZ, \rho\}.
\end{equation}
This can be implemented with jump operator $L = \sqrt{J} X$ along with postselected jump $\Tilde{L} = \sqrt{2h(I+Z)}$, and thus effective Hamiltonian $H_{\rm eff} = h Z$ up to a constant.
For this discussion, it is convenient to unravel the noise into ``strong'' jumps (i.e., complete bit flips): in each time interval $\Delta t$, there is a probability $J\Delta t$ of a complete spin flip ($\ket{0}\leftrightarrow\ket{1}$). Quantum trajectories therefore correspond to (classical) histories of the spin state: $\{s_t\}_{t=0}^{n}$, where each $s_t = \pm 1$ specifies whether the spin flips at the given time step. The spin state is given by the total number of flips (modulo 2) from the beginning to the current time.

\begin{figure}
    \centering
    \includegraphics[scale=1]{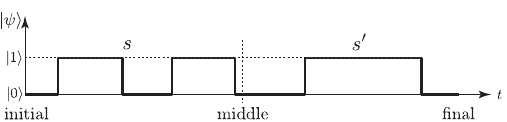}
    \caption{Illustration of the qubit trajectories discussed in Sec.~\ref{subsec: dimer example}. The trajectories $s$ and $s'$ can be combined into a single trajectory $\sigma$ with the initial and final states being $\ket{0}$. At each time step, the state $\ket{\psi_s}$ is proportional to either $\ket{0}$ or $\ket{1}$. At the middle point, we have $z(s)=z(s')$}
    \label{fig: ladder_trajectory}
\end{figure}

To obtain the desired expectation value, we must compute
\begin{equation}
    \langle Z_l\rangle 
    = \frac{\tr(\rho^2 Z)}{\tr(\rho^2)}
    = \frac{\sum_{s,s'}  \braket{\psi_{s'}|Z|\psi_s} \braket{\psi_s|\psi_{s'}} }{ \sum_{s,s'} |\braket{\psi_s|\psi_{s'}}|^2 }
\end{equation}
by sampling over trajectories $s,s'$. 
As discussed above, $\ket{\psi_s}$ is a classical state, proportional to either $\ket{0}$ or $\ket{1}$; let $\ket{\psi_s} = \sqrt{w(s)} \ket{z(s)}$. This gives $\braket{\psi_{s'}|\psi_s} = \sqrt{w(s) w(s')} \delta_{z(s),z(s')}$. It is convenient to view $s$ and $s'$ together as a single trajectory, $\sigma$, that starts and ends at $0$ (the initial state) and is continuous at the mid point as sketched in Fig. \ref{fig: ladder_trajectory}; we define also $w(\sigma) = w(s) w(s')$ and $z(\sigma) = z(s) = z(s')$ (value of the spin in the middle point). 
With this notation established, we may write 
\begin{equation}
    \langle Z_l\rangle 
    = \frac{\tr(\rho^2 Z)}{\tr(\rho^2)}
    = \frac{\sum_{\sigma} w(\sigma) z(\sigma)}{\sum_{\sigma} w(\sigma)} = \mathbb{E}_{\sigma\sim w} [z(\sigma)], 
    \label{eq: spin_ladder_exp}
\end{equation}
which is seen to be the average of the middle spin $z(\sigma)$ in a statistical-mechanical model with Boltzmann weights $w(\sigma)$. 

The Boltzmann weights have two factors: one associated to jumps,
\begin{equation}
    (J\Delta t)^{\sum_t (1-\sigma_t \sigma_{t+1})/2}
\end{equation}
(i.e. $J\Delta t$ if $\sigma_t\neq \sigma_{t+1}$, 1 otherwise), 
and another associated to ``sojourns'' in the $\ket{0}$ state (imaginary time evolution),  
\begin{equation} 
e^{h\Delta t \sum_t (1+\sigma_t)/2}.
\end{equation}
Overall this gives the Boltzmann weight of each trajectory pair $\sigma$ as
\begin{align}
    w(\sigma) = e^{-\frac{1}{2} \sum_t \log(J\Delta t) \sigma_t \sigma_{t+1} + h\Delta t \sigma_t + {\rm const.}} 
\end{align}
which corresponds to a 1D classical Ising model in a field. 
To efficiently sample the value of $z(\sigma)$, one must sample $\sigma$ configurations from this Ising model, taking into account the {\it global} effect of imaginary time evolution ($h$) along with the stochastic jumps ($J$). 

A na\"ive application of the prescription in Sec.~\ref{subsec: trajectory method} would independently sample configurations of $s$ and $s'$ based on the weight $\tilde{w}(s) \propto (J\Delta t)^{\sum_t (1-s_t s_{t+1})/2}$, which only takes into account the jumps and is evaluated locally in time (the Boltzmann weight decouples in the bond variables $s_t s_{t+1}$). This approach would give 
\begin{equation}
    \langle Z_l\rangle 
    = \frac{\tr(\rho^2 Z)}{\tr(\rho^2)}
    = \frac{\mathbb{E}_{\sigma \sim \tilde{w}} [e^{-h\Delta t\sum_t (1+\sigma_t)/2} z(\sigma)] }{\mathbb{E}_{\sigma \sim \tilde{w}} [e^{-h\Delta t\sum_t (1+\sigma_t)/2}]}
\end{equation}
The problem with this sampling of trajectories is that the arguments of the averages are broadly distributed (exponentially in $N$). 
As a consequence the averages will likely be dominated by configurations of $\sigma$ that are atypical with respect to the measure $\tilde{w}$, and the number of samples $\sigma\sim\tilde{w}$ needed for an accurate estimate will grow exponentially in $N$. 
This fact is closely related to the loss of normalization that we noted in Sec.~\ref{subsec: postselection} in the presence of postselection (represented here by the field $h$). 

This simple example illustrates the need for {\it importance sampling} of quantum trajectories in the presence of postselection, which we discuss next.

\subsection{Importance sampling \label{subsec: importance}}

Consider an observable $A_l$ in the virtual bilayer system, where $A$ is Pauli operator (for qubits) or a Majorana monomial (for fermions), such that $A = A^\dagger = A^{-1}$. 
Its expectation value maps onto a Renyi-2 correlator for the density matrix in the monolayer, 
\begin{equation}
{}_{lr}\! \bra{\Psi} A_l \ket{\Psi}_{lr} 
= \frac{\sbra{\rho} A\otimes I \sket{\rho}}{\sbraket{\rho}} 
= \frac{{\rm Tr}(\rho^2 A)}{{\rm Tr}(\rho^2)}.
\end{equation}
Next, we unravel the density matrix $\rho$ into un-normalized quantum trajectories $\{ \ket{\psi_{\vec s}} \}$ as in Eq.~\eqref{eq:trajectories_def}, such that $\rho = \sum_{\vec s} \ket{\psi_{\vec s}} \bra{\psi_{\vec s}}$.
For convenience, we assume the entries in $\vec s$ are bits---this can always be accomplished by splitting the noise channel associated to jumps $\{L_i\}$ into a product of noise channels, one for each jump $L_i$, with two Kraus operators.

In terms of these trajectory wavefunctions the R\'enyi-2 correlator reads
\begin{align}
    \frac{{\rm Tr}(\rho^2 A)}{{\rm Tr} (\rho^2)} 
    & = \frac{\sum_{\vec s, \vec s'} \braket{\psi_{\vec s}|\psi_{\vec s'}} \bra{\psi_{\vec s'}} A \ket{\psi_{\vec s}} }
    {\sum_{\vec s, \vec s'} |\braket{\psi_{\vec s}|\psi_{\vec s'}}|^2} \nonumber \\
    & = \sum_{\vec s, \vec s'} 
    \left( \frac{ |\braket{\psi_{\vec s}|\psi_{\vec s'}}|^2}{\sum_{\vec r, \vec r'} | \braket{\psi_{\vec r}|\psi_{\vec r'}}|^2 } \right) 
    \frac{ \bra{\psi_{\vec s'}} A \ket{\psi_{\vec s}} }{ \braket{\psi_{\vec s'} |\psi_{\vec s}} } \nonumber \\
    & = \sum_{\vec s, \vec s'} p(\vec s, \vec s') \frac{ \bra{\psi_{\vec s'}} A \ket{\psi_{\vec s}} }{ \braket{\psi_{\vec s'} |\psi_{\vec s}} } \label{eq:importance_intralayer}.
\end{align}
This is an average of the function $\bra{\psi_{\vec s'}} A \ket{\psi_{\vec s}} / \braket{\psi_{\vec s'} |\psi_{\vec s}}$ over trajectory pairs $(\vec s, \vec s')$ drawn according to the probability distribution $p(\vec s,\vec s') \propto |\braket{\psi_{\vec s}|\psi_{\vec s'}}|^2$.  
At this point, the correlator can be estimated by Monte Carlo sampling trajectory pairs according to $p(\vec s, \vec s')$, for example by Metropolis sampling:
\begin{enumerate}
    \item Draw random bitstrings $\vec s, \vec s'$ and compute the corresponding un-normalized trajectory wavefunctions $\ket{\psi_{\vec s}}$, $\ket{\psi_{\vec s'}}$;
    \item Flip a randomly chosen bit in $(\vec s, \vec s')$ (i.e., change the choice of jump at one space-time point) to obtain a new pair $(\vec r, \vec r')$, and compute $\ket{\psi_{\vec r}}$, $\ket{\psi_{\vec r'}}$; 
    \item Compute the ratio of squared overlaps $f \equiv |\braket{\psi_{\vec r}|\psi_{\vec r'}}|^2 / |\braket{\psi_{\vec s}|\psi_{\vec s'}}|^2$;
    \item Accept the update with probability $\min(1,f)$, repeat. 
\end{enumerate}
This Markov chain equilibrates to the desired distribution $p(\vec s, \vec s')$. After equibration, we draw samples $(\vec s,\vec s')$ from the Markov chain, compute the ratio $\bra{\psi_{\vec s'}} A \ket{\psi_{\vec s}} / \braket{\psi_{\vec s'} |\psi_{\vec s}}$, and average to obtain the R\'enyi correlator. 

A possible issue with this approach is that the argument of the average is not bounded (one can have, e.g., $\braket{\psi_{\vec s} | \psi_{\vec s'}} = 0$ with $\braket{\psi_{\vec s} | A| \psi_{\vec s'}} \neq 0$), so the Monte Carlo sampling is not guaranteed to converge efficiently in the number of samples. 
To address this issue we show that the random variable to be averaged has bounded variance. 
First, let us introduce the shorthand notation $A_{\vec s, \vec s'} \equiv \bra{\psi_{\vec s}} A \ket{\psi_{\vec s'}} $, $I_{\vec s,\vec s'} \equiv \braket{\psi_{\vec s} | \psi_{\vec s'}}$.
By Eq.~\eqref{eq:importance_intralayer} we have the mean
\begin{equation}
    \mathop{\mathbb{E}}_{(\vec s, \vec s') \sim p} \left[ \frac{A_{\vec s,\vec s'}}{I_{\vec s,\vec s'}} \right] = \frac{{\rm Tr}(\rho^2 A)}{{\rm Tr}(\rho^2)}.
\end{equation}
The variance is
\begin{align}
    \mathop{{\rm var}}_{ (\vec s, \vec s') \sim p } \left[\frac{A_{\vec s,\vec s'}}{I_{\vec s,\vec s'}}\right] 
    & = \mathop{\mathbb{E}}_{(\vec s, \vec s') \sim p} \left[\left|\frac{A_{\vec s,\vec s'}}{I_{\vec s,\vec s'}}\right|^2 \right] 
    - \left| \mathop{\mathbb{E}}_{(\vec s, \vec s') \sim p} \left[\frac{A_{\vec s,\vec s'}}{I_{\vec s,\vec s'}}\right]\right|^2 \nonumber \\
    & = \sum_{\vec s, \vec s'} p(\vec s, \vec s') \frac{|A_{\vec s,\vec s'}|^2}{|I_{\vec s, \vec s'}|^2} - \left(\frac{{\rm Tr}(\rho^2 A)}{{\rm Tr}(\rho^2)} \right)^2 \nonumber \\
    & = \frac{ \sum_{\vec s, \vec s'} |A_{\vec s,\vec s'}|^2 }{ \sum_{\vec s, \vec s'} |I_{\vec s,\vec s'}|^2 } - \left(\frac{{\rm Tr}(\rho^2 A)}{{\rm Tr}(\rho^2)} \right)^2 \nonumber \\
    & = \frac{{\rm Tr}(\rho A \rho A)}{{\rm Tr}(\rho^2)} - \left(\frac{{\rm Tr}(\rho^2 A)}{{\rm Tr}(\rho^2)} \right)^2 . 
\end{align}
This is bounded above as 
\begin{align}
    \mathop{{\rm var}}_{ (\vec s, \vec s') \sim p } \left[\frac{A_{\vec s,\vec s'}}{I_{\vec s,\vec s'}}\right] 
    & \leq \frac{{\rm Tr}(\rho A \rho A)}{{\rm Tr}(\rho^2)} 
    = {}_{lr}\!\bra{\Psi} A_l \bar{A}_r \ket{\Psi}_{lr} \leq 1.
\end{align}
Given that the variance is finite, we can use for example the median-of-means estimator \cite{doi:10.1137/1027074} to obtain guaranteed convergence within additive error $\varepsilon$ from $O(1/\varepsilon^2)$ samples with high probability.

Consider next an inter-layer operator of the form $A_l \bar{A}_r$ in the virtual bilayer state, corresponding to the Renyi-2 correlator ${\rm Tr}(\rho A \rho A) / {\rm Tr}(\rho^2)$ on the physical density matrix. 
Assuming ${\rm Tr}(\rho^2 A)\neq 0$, we may write 
\begin{align}
    \frac{{\rm Tr}(\rho A \rho A)}{{\rm Tr} (\rho^2)} 
    = \frac{{\rm Tr}(\rho^2 A)}{{\rm Tr} (\rho^2)} \frac{{\rm Tr}(\rho A \rho A)}{{\rm Tr} (\rho^2 A)}
    \label{eq: inter_operator_value}
\end{align}
where the first factor is the one we already discussed above. As for the second factor, we have  
\begin{align}
    \frac{{\rm Tr}(\rho A \rho A)}{{\rm Tr} (\rho^2 A)} 
    & = \frac 
    {\sum_{\vec s, \vec s'} |A_{\vec s, \vec s'}|^2 }
    {\sum_{\vec s, \vec s'} I_{\vec s, \vec s'} A_{\vec s',\vec s}} \nonumber \\
    & = \left( \frac 
    {\sum_{\vec s, \vec s'} |A_{\vec s, \vec s'}|^2 ( I_{\vec s, \vec s'} / A_{\vec s,\vec s'} )} 
    {\sum_{\vec s, \vec s'} |A_{\vec s, \vec s'}|^2 } \right)^{-1} \nonumber \\
    & = \left( \mathop{\mathbb{E}}_{(\vec s,\vec s')\sim q} [ I_{\vec s, \vec s'} / A_{\vec s,\vec s'} ]\right)^{-1}
\end{align}
in terms of a probability distribution 
\begin{equation} 
q(\vec s, \vec s') = \frac{ |A_{\vec s, \vec s'}|^2}{ \sum_{\vec r, \vec r'} |A_{\vec r, \vec r'}|^2}. 
\end{equation}
Once again, the argument of the average is unbounded (one can have $A_{\vec s,\vec s'} = 0$ with $I_{\vec s,\vec s'} \neq 0$) but we can bound its variance. 
We have 
\begin{align}
    \mathop{{\rm var}}_{ (\vec s, \vec s') \sim q } \left[\frac{I_{\vec s,\vec s'}}{A_{\vec s,\vec s'}}\right] 
    & \leq \sum_{\vec s, \vec s'} q(\vec s,\vec s') \frac{|I_{\vec s,\vec s'}|^2}{|A_{\vec s,\vec s'}|^2} \nonumber \\
    & \leq \frac
    { \sum_{\vec s, \vec s'} |I_{\vec s,\vec s'}|^2 }
    { \sum_{\vec s, \vec s'} |A_{\vec s,\vec s'}|^2 }
    = \frac{{\rm Tr}(\rho^2)}{{\rm Tr}(\rho A \rho A)}.  \label{eq:variance_bound_interlayer}
\end{align}
This bound is the inverse of the quantity to be estimated.
Provided that both ${\rm Tr}(\rho A \rho A) / {\rm Tr}(\rho^2)$ and ${\rm Tr}(\rho^2 A) / {\rm Tr}(\rho^2)$ are gapped away from zero, the estimation is sample-efficient. 

To summarize, we can express correlators in the virtual bilayer system in terms of quantum trajectory averages on the open monolayer system, and we can implement importance-sampling methods that make the sampling efficient. 
We have
\begin{align}
    \bra{\Psi} A_l \ket{\Psi}
    & = \frac{{\rm Tr}(\rho^2 A)}{{\rm Tr} (\rho^2)} 
    =  \mathop{\mathbb{E}}_{\vec s, \vec s' \sim p} [ A_{\vec s, \vec s'} / I_{\vec s, \vec s'}] 
\end{align}
for intra-layer operators and
\begin{align}
    \bra{\Psi} A_l \bar{A}_r \ket{\Psi}
    & = \frac{{\rm Tr}(\rho A \rho A)}{{\rm Tr} (\rho^2)} 
    = \frac
    { \mathbb{E}_{\vec s, \vec s' \sim p} [ A_{\vec s, \vec s'} / I_{\vec s, \vec s'}]}
    {\mathbb{E}_{\vec s, \vec s' \sim q} [I_{\vec s, \vec s'} / A_{\vec s, \vec s'}]} 
\end{align}
for inter-layer operators obeying antiunitary layer-exchange symmetry, in terms of probability distributions $p \propto |I_{\vec s, \vec s'}|^2$ and $q \propto |A_{\vec s, \vec s'}|^2$ over pairs of trajectories.

\section{Connection to auxiliary-field quantum Monte Carlo \label{sec: afqmc}}

The mapping presented in Sec.~\ref{sec: Setup} and the quantum trajectory sampling method of Sec.~\ref{sec: quantum trajectory} can be combined into a numerical method to simulate low-energy states of bilayer quantum Hamiltonians. By simulating the dynamics of monolayer systems, the system size is effectively reduced by a factor of 2 at the expense of some trajectory averaging (which can be made efficient by importance sampling). 
This leads, asymptotically, to a quadratic reduction in the problem's computational complexity. 

A greater reduction in complexity is achieved when the quantum trajectory unraveling produces non-interacting fermionic dynamics, which can be simulated efficiently (with time and memory scaling polynomially in $N$). 
Below we show that this approach is closely related to auxiliary-field quantum Monte Carlo (AFQMC) \cite{PhysRevD.24.2278,PhysRevB.55.7464,PhysRevLett.90.136401,lee2022yearsauxiliaryfieldquantummonte}, a standard method in the simulation of strongly-correlated quantum matter. In particular, we show that known criteria for evading the ``sign problem'' in AFQMC acquire a physical interpretation in terms of constraints on the monolayer system's dynamics.

We briefly review AFQMC with an emphasis on aspects relevant to this work. Our discussion is purely qualitative, and we refer to Refs.~\cite{10.1063/5.0031024,lee2022yearsauxiliaryfieldquantummonte} for details. 

AFQMC is a method to estimate observables in ground states or low-temperature states of certain interacting fermionic systems. 
The method is based on approximating the ground state wavefunction $\ket{\Psi_\text{GS}}$ by imaginary time evolution of an initial state $\ket{\Psi_\text{I}}$:
\begin{equation}
    \ket{\Psi_\text{GS}} \propto \lim_{\beta\rightarrow\infty} e^{-\beta \H}\ket{\Psi_\text{I}} = \lim_{n\rightarrow\infty}(e^{-\Delta \tau \H})^n\ket{\Psi_\text{I}}.
\end{equation}
The key idea is to simplify this interacting many-body calculation by decoupling the interactions into some channel, e.g., the density channel. 
This can be accomplished with a Hubbard-Stratonovich transformation\footnote{Actual computational implementations differ in the specifics of how to decouple interactions.},
\begin{equation}
    e^{-V(\bar\psi \psi)^2} \propto \int d\varphi e^{i\varphi \bar\psi \psi} e^{-\frac{1}{4V} \varphi^2},
\end{equation}
where $\psi$ represents the fermion field and $\varphi$ the auxiliary field---a classical variable that is being averaged over and couples to the fermion density $\bar\psi \psi$ like an applied field. 
Iterating this Hubbard-Stratonovich transformation at each point in space and time yields 
\begin{equation}
    e^{-\Delta \tau \H} = \int D\vec\varphi~ p(\vec\varphi) e^{-\int d\tau (\bar\psi_\tau h \psi_\tau -i \bar\psi_\tau \psi_\tau \varphi_\tau)},
    \label{eq: HS}
\end{equation}
where $\vec\varphi = \{\varphi_\tau\}$ is a configuration of auxiliary fields, $D\vec\varphi$ represents integration over these configurations, $p(\vec\varphi)$ is the Gaussian distribution, and $h$ is an unspecified non-interacting fermionic Hamiltonian. We suppress the spatial dependence in our notation for simplicity. 
The idea of AFQMC is to randomly sample configurations of the auxiliary field, efficiently simulate the resulting free-fermion dynamics, and thus obtain an ensemble of free-fermion wavefunctions (Slater determinants) that can be used to estimate observables in the original interacting problem. 

The main limitation of quantum Monte Carlo approaches including AFQMC is that, while they yield the correct result on average, statistical fluctuations are not always controlled. This is known as the ``sign problem'' (or ``phase problem''), describing the fact that physical quantities are given by sums of many large terms with variable sign (or phase) that cancel out only on average. On the contrary, in models without a sign problem, one gets sums involving positive terms only, where large fluctuations and cancellations are not possible and sampling is thus efficient.  
For instance, in the repulsive Hubbard model at half-filling, particle-hole symmetry implies that the spin-up and spin-down propagators are complex conjugates of each other, making the total weight an absolute value squared hence non-negative. However, in more generic settings---such as the Hubbard model away from half-filling---this symmetry no longer holds, and a sign problem typically emerges. To address this, advanced techniques like constrained-path AFQMC \cite{PhysRevB.55.7464} and phaseless AFQMC \cite{PhysRevLett.90.136401} can be employed to enhance computational efficiency. We will not discuss these approaches in this work. 

Eq.~\eqref{eq: HS} has some notable similarities with our approach: it mixes imaginary time evolution (the term $e^{-\int \bar\psi h \psi}$) with noisy real-time evolution (the term $e^{i\int \bar\psi\psi\varphi}$, with $\varphi$ a classical variable to be averaged over). 
Comparing with the quantum trajectory method, the auxiliary field $\varphi$ corresponds to the random variables $dw_i$ in Eq.~\eqref{eq: unravel pure states}, the operator that couples to $\varphi$ (the density $\bar\psi\psi$ in our example) corresponds to the jump operator $L_i$, and the Hamiltonian $h$ corresponds to $H_{\rm eff}$. Thus we may interpret AFQMC as a sampling of quantum trajectories for the dynamics of a monitored free-fermionic system subject do density dephasing. 

Let us now specialize to Hubbard-like systems, with Hamiltonians
\begin{equation}
\mathcal{H} = -\sum_{i,j,\sigma} t_{ij} c^\dagger_{i,\sigma} c_{j,\sigma} + U \sum_i \left( n_{i,\uparrow} - \frac{1}{2} \right) \left( n_{i,\downarrow} - \frac{1}{2} \right) \label{eq:hubbard_like}
\end{equation}
comprising two species of fermions $\sigma = \uparrow, \downarrow$ (typically taken to be spin) with intra-species hopping and inter-species interaction. These models may or may not present a sign problem for AFQMC depending on Hamiltonian parameters, lattice structure, and electron filling for the two species. 
In particular, the model is sign-problem-free when interactions are repulsive ($U>0$), the underlying lattice (specified by the $t_{ij}$ hopping amplitudes) is bipartite, and the system is at half filling (i.e. there is on average one electron per site, $\nu_\uparrow + \nu_\downarrow = 1$). These conditions lead to an antiunitary symmetry between the species $\sigma = \uparrow, \downarrow$, causing the relevant amplitudes in the calculation to come in complex-conjugate pairs and thus eliminating sign or phase fluctuations.  

While these criteria are straightforward to verify mathematically, they lack an obvious physical interpretation. Our mapping to open-system dynamics provides such an interpretation by identifying the two species $\sigma = \uparrow,\downarrow$ with the two sides of a density matrix, the ``ket'' Hilbert space $l$ and the ``bra'' Hilbert space $r$. 
Here we review the criteria for sign-free AFQMC in Hubbard models and relate each one of them to our mapping (Sec.~\ref{sec: Setup}): 
\begin{itemize}
    \item {\it Antiunitary layer-exchange symmetry}. This requirement stems from the fact that the two layers $l$, $r$ originate from the ``ket'' and ``bra'' sides of a single physical density matrix, whose evolution is not independent but must be related by time reversal. 
    \item 
    {\it Sign of the couplings.} The inter-layer couplings $J_i$ in our mapping, $J_i O_{i,l} \bar{O}_{i,r}$, are non-negative: they are the squares of jump operator amplitudes. If $U>0$ (repulsive interactions), this requires $O_i\equiv n_i-1/2$ to be odd under the antiunitry map, $\bar{O}_i = -O_i$. This singles out particle-hole symmetry, $\P$, as the antiunitary map to use. 
    If $U<0$ (attractive interactions), the same reasoning yields time-reversal $\T$.
    \item {\it Half filling}. Assuming $U>0$, the ``vectorization'' of the density matrix $\rho = \ket{\psi} \bra{\psi} \mapsto \sket{\rho}_{lr} = \ket{\psi}_l \ket{\bar{\psi}}_r$ is implemented by particle-hole symmetry, $\P$, as described above. So if $\ket{\psi}$ is a state with filling $\nu$, then $\ket{\bar{\psi}}$ has the complementary filling $1-\nu$. Total filling in the two layers is thus exactly pinned to $\nu_l + \nu_r = 1$ (opposite fluctuations within the layers are allowed). 
    \item {\it Bipartite lattice}. Assuming $U>0$, we have $\bar{H} = \P H \P^{-1}$, which changes the sign of the intra-layer hopping Hamiltonian $H = \sum_{ij} t_{ij} c_i^\dagger c_j$: $\bar{H} = -H$ (assuming $t_{ij}$ real). This relative sign between the hopping terms of each species does not appear in Eq.~\eqref{eq:hubbard_like}. 
    However, if the lattice is bipartite, this sign is physically irrelevant: it can be eliminated with a unitary transformation on $r$, $V_B = \prod_{j\in B} (-1)^{n_{j,r}}$ with $B$ a sublattice.
\end{itemize}

Thus we see that the criteria for sign-free AFQMC reflect constraints on the structure of density matrices---specifically Hermiticity and constraints on particle numbers between ``ket'' and ``bra'' components.

\section{Example: 1D Ashkin-Teller model \label{sec: examples}}

\begin{figure}
    \centering
    \includegraphics[scale=1.1]{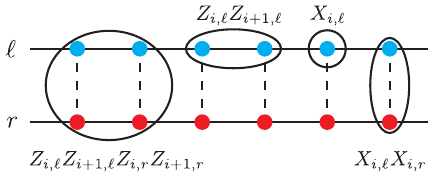}
    \caption{The Hamiltonian of 1D quantum Ashkin-Teller model consists of two copies of transverse-field Ising models represented as blue and red dots respectively, with an interlayer coupling tuned by $\lambda_J$ and $\lambda_h$ represented as dashed rungs.}
    \label{fig: 1D_AT}
\end{figure}

In this section, we apply the numerical method based on quantum trajectories with importance sampling to the 1D quantum Ashkin-Teller model. 
This model naturally fits into our framework and is illustrated in Fig. \ref{fig: 1D_AT}. The Hamiltonian is given by
\begin{align}
    \H_\text{AT} = & J\sum_i (Z_{i,l}Z_{i+1,l} + Z_{i,r} Z_{i+1,r}-\lambda_J Z_{i,l}Z_{i+1,l}Z_{i,r}Z_{i+1,r}) \nonumber \\
    & + h \sum_i (X_{i,l} - X_{i,r} + \lambda_h X_{i,l} X_{i,r}).
\end{align}
Note the minus sign on the $X_{i,r}$ term can be absorbed into $\lambda_h$ by a unitary transformation $\prod_i Z_{i,r}$ on the right layer.

This Hamiltonian can be realized by an open system with non-monitored jumps
\begin{align}
    L_{1,i} = \sqrt{J\lambda_J} Z_i Z_{i+1},\quad L_{2,i} = \sqrt{h\lambda_h} X_i,
\end{align}
and postselected jumps
\begin{align}
    \tilde{L}_{1,i} = \sqrt{J}(1 + Z_{i}Z_{i+1}),\quad \tilde{L}_{2,i} = \sqrt{h}(1 + X_i).
\end{align}
giving the effective intralayer Hamiltonian (up to additive constants)
\begin{equation}
    H_\text{eff} = J\sum_{i}Z_{i}Z_{i+1} + h\sum_i X_i,
\end{equation}
a transverse-field Ising model.

We calculate the intralayer correlator 
\begin{equation}
    C^{(1)} 
    = \bra{\Psi} Z_{1,l} Z_{2,l} \ket{\Psi}
    = \frac{{\rm Tr}(\rho^2 Z_1 Z_2) }{{\rm Tr}(\rho^2) }
\end{equation}
and the interlayer correlator
\begin{equation}
    C^{(2)} 
    = \bra{\Psi} Z_{1,l} Z_{2,l} Z_{1,r} Z_{2,r} \ket{\Psi}
    = \frac{{\rm Tr}(\rho Z_1 Z_2 \rho Z_1 Z_2) }{{\rm Tr}(\rho^2) }
\end{equation}
on a system of size $L=8$ with open boundary conditions. 
We carry out exact simulation of the bilayer system (still feasible for $2L = 16$ qubits) as well as simulation by importance-sampled quantum trajectories as discussed in Sec. \ref{subsec: importance}. 
Starting from the initial state $\rho = (\ket{1}\bra{1})^{\otimes 2} \otimes (\ket{0}\bra{0})^{\otimes 6}$, we compute the time evolution in the open monolayer system and compare it with the thermal evolution ($e^{-\beta\H}$) of the bilayer. The latter is computed by Krylov time evolution, trotterized in the same way as the trajectories so that the two should agree exactly in the limit of infinitely many trajectory samples.
Parameter values are chosen to be $J=1$, $h=0.3$, while $\lambda_J=\lambda_h$ is varied across $0.1,~ 0.5,$ and $1$. All three cases are computed by averaging $128$ runs of $2\times 10^5$ Monte Carlo updates (each one started from an independently sampled random trajectory). The inverse temperature $\beta$ is identified with the evolution time $t$, and the dynamics is carried out up to $\beta = 2$ with the Trotterization $\Delta \beta =0.1$. 

\begin{figure}
    \centering
    \includegraphics[scale=0.55]{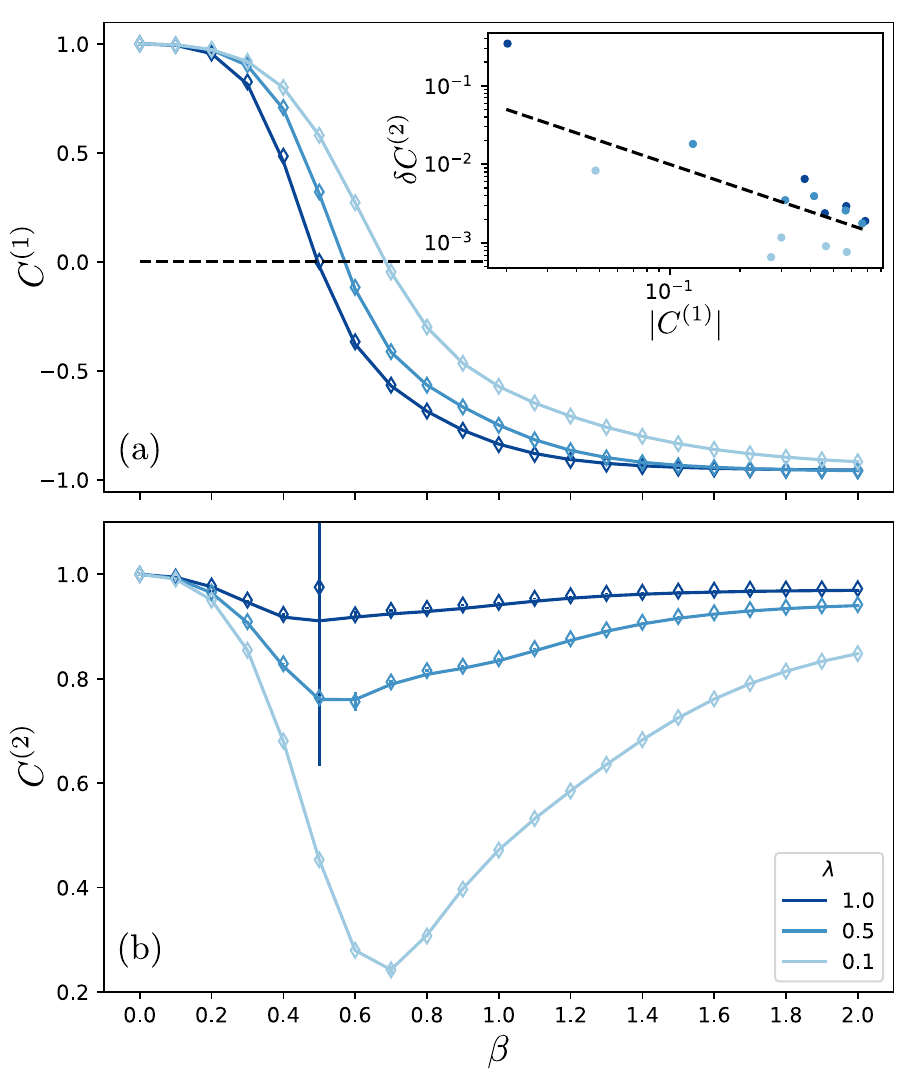}
    \caption{Numerical benchmark of importance-sampled quantum trajectory method. We simulate a 1D quantum Ashkin-Teller model with $L = 8$, open boundary conditions, and parameters $J=1$, $h=0.3$, $\lambda_J = \lambda_h = \lambda$ (see legend). 
    (a) Intralayer correlator $C^{(1)} = \langle Z_{1,l} Z_{2,l}\rangle$ and
    (b) interlayer correlator $C^{(2)} = \langle Z_{1,l} Z_{2,l} Z_{1,r} Z_{2,r}\rangle$ vs inverse temperature $\beta$.
    Empty diamonds show results of quantum trajectory simulation, averaging 128 runs with $2\times 10^5$ Metropolis updates each. 
    Solid lines show exact Krylov imaginary time evolution under $\H$ (both approaches are Trotterized with $\Delta\beta = 0.1$). 
    Error bars are mostly invisible, with the exception of $\beta=0.5$ and $\lambda = 1$ which displays very large uncertainty. This is explained by the near-zero value of $C^{(1)}$ at the same point (see main text). Inset: statistical uncertainties on $C^{(2)}$ as a function of $|C^{(1)}|$, the dashed line shows inverse proportionality for reference.
    }
    \label{fig: 1D_AT_ZZ}
\end{figure}

Results for the intralayer correlator $C^{(1)}$ are shown in Fig.~\ref{fig: 1D_AT_ZZ}(a) and for the interlayer correlator $C^{(2)}$ in Fig.~\ref{fig: 1D_AT_ZZ}(b). 
The diamond markers represent the importance-sampled quantum trajectory results, while the solid lines denote exact simulation results from the Krylov method with bilayer Hamiltonian $\H_\text{AT}$.
Error bars for each point come from statistical fluctuations between quantum trajectories, computed using the method of batch means~\cite{10.1063/1.457480,doi:10.1287/ijoc.9.3.296,32b1381e-f1dc-38d2-9726-9fc84485ef1a}. Most of them are on the order of $10^{-3}$ to $10^{-4}$ and are therefore invisible in Fig. \ref{fig: 1D_AT_ZZ}. However, an exception happens for the interlayer correlator $C^{(2)}$ at $\beta=0.5$, $\lambda=1$, where statistical fluctuations become extremely large such that the value cannot be estimated accurately with the given number of Monte Carlo steps. 
The issue stems from the fact that the intralayer correlator $C^{(1)}$ is very close to zero at that point, as can be seen in Fig. \ref{fig: 1D_AT_ZZ}(a). The interlayer correlator $C^{(2)}$ is calculated using Eq. \eqref{eq: inter_operator_value}, where the expectation value is expressed as a ratio of two quantities to be evaluated separately; the denominator is $C^{(1)}$. Thus, even though the absolute variance of $C^{(1)}$ is bounded, when $C^{(1)} \approx 0$ the {\it relative} error diverges resulting in the observed behavior.
A similar situation occurs at $\beta = 0.7$ and $\lambda=0.1$, where the numerator is also nearly zero. However, because the interlayer coupling is weaker in this case, fluctuations across trajectories are overall smaller and the number of samples we employ is sufficient to yield an accurate result. 

This benchmark confirms the correctness of our mapping and importance sampling scheme, while also illustrating possible issues due to zeros of correlation functions. Another possible issue, which we do not encounter in this case, is the vanishing of interlayer correlators leading to unbounded variance [Eq.~\eqref{eq:variance_bound_interlayer}]. 
We note also that, in this model, the trajectories reduce to free fermions (Ising model), so it would be possible to scale the simulation to much larger system sizes by fermionic Gaussian state methods~\cite{10.5555/2011637.2011640,Prosen_2008,10.21468/SciPostPhysLectNotes.54}. For the purpose of this work, we ignored the free-fermion nature of the model and used full many-body wavefunction simulations, but the implementation of free-fermion trajectory simulations (recovering AFQMC) is an interesting direction for future work. 

\section{Summary and discussion} \label{sec:discussion}

We studied the connection between mixed states and pure states in a doubled system. This connection is typically employed to understand open systems in terms of isolated ones, for which more theoretical tools are available. Here we took the opposite route, proposing to map bilayer Hamiltonians onto open and monitored dynamics of corresponding monolayers. 
We derived criteria under which this mapping is possible: (i) an antiunitary layer-exchange symmetry, capturing the relationship between ``ket'' and ``bra'' Hilbert spaces, and (ii) a sign constraint on interlayer couplings, arising from the positivity of density matrices. 
Thermal states of the bilayer Hamiltonian are identified with dynamical states of the monolayer at time $t = \beta$; in particular ground states of the former correspond to steady states of the latter. 
This connects quantum phase transitions in bilayer systems with non-equilibrium phase transitions in open and monitored systems, which will be interesting to explore more deeply in future work. 

Our approach differs from other related recent works \cite{yan2024dissipative,pjs3-14cc, PRXQuantum.5.020332, lu2024bilayerconstructionmixedstate, PhysRevB.110.L241105, PhysRevB.111.054106} by the inclusion of postselection, which allows for the simulation of a wide and physically interesting class of bilayer Hamiltonians [all those meeting criteria (i) and (ii) above]. When interlayer and intralayer couplings are ``locked'' in a specific way, the mapping results in purely dissipative dynamics, without postselection, in the monolayer system; allowing for some amount of postselection makes it possible to explore more choices of couplings, and correspondingly a wider range of physical phenomena.  

We also presented a computational method based on this mapping, where low-temperature observables in bilayer systems are estimated from quantum trajectory evolution in monolayer systems, potentially resulting in quadratic reductions in computational cost (i.e., doubling of achievable system sizes). 
The overhead from sampling of multiple quantum trajectories can be controlled by a suitable importance sampling scheme, which gives bounded statistical variance for many observables even in the presence of postselection.
When the quantum trajectories describe non-interacting fermions, the method becomes equivalent to AFQMC, with the trajectory label playing the role of the auxiliary field configuration. Notably, this gives a physically intuitive interpretation of the Monte Carlo sign-free criteria for bilayer Hamiltonians such as the Hubbard model.

Many interesting bilayer systems fall under the assumptions of our mapping, including the Fermi-Hubbard model and various quantum Hall bilayers hosting exciton condensation transitions~\cite{eisenstein2014qhbilayers}, which may be interpreted as SWSSB of the $U(1)$ charge conservation symmetry~\cite{PhysRevB.110.155150,PRXQuantum.6.010344,pjs3-14cc}. 
Furthermore, recent advances in moir\'e physics \cite{doi:10.1073/pnas.1108174108,Cao20181,Cao20182} have demonstrated that twisted bilayer structures can exhibit a wide range of exotic phenomena, such as Skyrmion superconductivity \cite{doi:10.1126/sciadv.abf5299,PhysRevB.106.035421}, arising from the coexistence of nearly-flat bands with opposite nontrivial Chern numbers, e.g., $C = \pm 1$. It would be interesting to realize these models from open and monitored dynamics of a single topological band, $C = +1$, with the time-reversed $C = -1$ band emerging in the virtual layer. 
Understanding how these exotic bilayer phenomena may appear in monolayer density matrices is an interesting question for future work. 

\begin{acknowledgements}
We thank Peize Ding, Bishoy Kousa, Jose Manuel Torres-Lopez, and Zihao Qi for helpful discussions and comments on the manuscript. 
Numerical simulations were performed in part on HPC resources provided by the Texas Advanced Computing Center (TACC) at the University of Texas at Austin.
\end{acknowledgements}

{\it Data availability statement.}
The data that support the findings of this article are openly available~\cite{data_repository}.

\bibliography{refs}
\end{document}